\documentclass[conference]{IEEEtran}

\usepackage{amsfonts}
\usepackage{epsfig} 
\usepackage{amsmath,amsthm} 
\usepackage{amssymb}  
\usepackage{graphicx}

\newtheorem{theorem}{Theorem}

\newcommand{\sr}{\stackrel}

\newcommand{\tri}{\sr{\triangle}{=}}

\newcommand{\noi}{\noindent}

\newcommand{\be}{\begin{equation}}
\newcommand{\ee}{\end{equation}}
\newcommand{\bea}{\begin{eqnarray}}
\newcommand{\eea}{\end{eqnarray}}
\newcommand{\bes}{\begin{eqnarray*}}
\newcommand{\ees}{\end{eqnarray*}}

\newcommand{\bfi}{\begin{figure}}
\newcommand{\bfit}{\begin{figure}[t]}
\newcommand{\bfib}{\begin{figure}[b]}
\newcommand{\bfih}{\begin{figure}[h]}
\newcommand{\bfip}{\begin{figure}[p]}
\newcommand{\efi}{\end{figure}}

\newcommand{\bi}{\begin{itemize}}
\newcommand{\ei}{\end{itemize}}
\newcommand{\ben}{\begin{enumerate}}
\newcommand{\een}{\end{enumerate}}

\ifCLASSINFOpdf
\else
\fi

\renewcommand{\IEEEQED}{\IEEEQEDclosed}

\begin{document}
%
\title{Directed Information on Abstract Spaces: Properties and Extremum Problems}

\author{\IEEEauthorblockN{\bf Charalambos D. Charalambous and Photios A. Stavrou}
\IEEEauthorblockA{ECE Department, University of Cyprus, Nicosia, Cyprus\\
e-mail:\{\it~chadcha, stavrou.fotios\}@ucy.ac.cy}}

\maketitle

\begin{abstract}
This paper describes a framework in which directed information is defined on abstract spaces. The framework is employed to derive properties of directed information such as convexity, concavity, lower semicontinuity, by using the topology of weak convergence of probability measures on Polish spaces. Two extremum problems of directed information related to capacity of channels with memory and feedback, and non-anticipative and sequential rate distortion are analyzed showing existence of maximizing and minimizing distributions, respectively. 
\end{abstract}

\IEEEpeerreviewmaketitle

\section{Introduction}
Directed information  from a sequence of Random Variables (RV's) $X^n\tri\{X_0,X_1,\ldots,X_n\}\in{\cal X}_{0,n}\tri\times_{i=0}^n{\cal X}_i$, to another sequence $Y^n\tri\{Y_0,Y_1,\ldots,Y_n\}\in{\cal Y}_{0,n}\tri\times_{i=0}^n{\cal Y}_i$ is often defined via \cite{marko,massey90}\footnote{Unless otherwise, integrals with respect to measures are over the spaces on which these are defined.}
\begin{align}
&I(X^n\rightarrow{Y}^n)\tri\sum_{i=0}^{n}I(X^i;Y_i|Y^{i-1})\label{equation1a}\\
&=\sum_{i=0}^n\int\log\bigg(\frac{P_{Y_i|Y^{i-1},X^i}(dy_i|y^{i-1},x^i)}{P_{Y_i|Y^{i-1}}(dy_i|y^{i-1})}\bigg){P}_{X^i,Y^i}(dx^i,dy^i)\label{equation1b}\\
&\equiv\mathbb{I}_{X^n\rightarrow{Y}^n}(P_{X_i|X^{i-1},Y^{i-1}},P_{Y_i|Y^{i-1},X^i}:i=0,\ldots,n)\label{equation1c}
\end{align}
Since the joint distribution in (\ref{equation1b}) is decomposed via $P_{X^i,Y^i}(dx^i,dy^i)=\otimes_{j=0}^i{P}_{X_j|X^{j-1},Y^{j-1}}(dx_j|x^{j-1},y^{j-1})\otimes{P}_{Y_j|Y^{j-1},X^j}(dy_j|y^{j-1},x^j)$, the notation $\mathbb{I}_{X^n\rightarrow{Y}^n}(\cdot,\cdot)$ denotes the functional dependence on two collections of non-anticipative or causal conditional distributions $\{P_{X_i|X^{i-1},Y^{i-1}}(\cdot|\cdot,\cdot),~P_{Y_i|Y^{i-1},X^i}(\cdot|\cdot,\cdot)~:~i=0,1,\ldots,n\}$.\\
In information theory, directed information (\ref{equation1a})-(\ref{equation1c}) or its variants are used to characterize capacity of channels with memory and feedback  \cite{tatikonda2009,tatikonda2000,chen-berger2005}, lossy data compression with feedforward information at the decoder \cite{venkataramanan2008}, lossy data compression of sequential codes \cite{tatikonda2000}, lossy data compression of non-anticipative codes \cite{stavrou-charalambous-kourtellaris2012}, and capacity of networks such as the two-way channel, multiple access channel \cite{kramer1998,kramer2003}, etc. The previous references derive coding theorems based on  $a)$ stationary ergodic processes $\{(X_i,Y_i)\}_{i=0}^{\infty}$, $b)$ Dobrushin's stability of the information density $\log\otimes_{i=0}^n\frac{P_{Y_i|Y^{i-1},X^i}(dy_i|y^{i-1},x^i)}{P_{Y_i|Y^{i-1}}(dy_i|y^{i-1})}$, and $c)$ via information spectrum methods \cite{han93}.\\
\noi{\bf Capacity of Channels with Feedback.} Based on $a)$ or $b)$ the operational definition of channels with memory and feedback is given by 
\begin{align}
C^{f}(P)\tri\lim_{n\rightarrow\infty}\sup_{\overleftarrow{P}_{X^n|Y^{n-1}}(\cdot|\cdot)\in\overleftarrow{\cal{P}}(P)}\frac{1}{n+1}I(X^n\rightarrow{Y^n})\label{equation1d}
\end{align}
where $\overleftarrow{\cal{P}}(P)$ denotes the power constraint set, and 
\begin{align}
 \overleftarrow{P}_{X^n|Y^{n-1}}(dx^n|dy^{n-1})\tri\otimes_{i=0}^n{P}_{X_i|X^{i-1},Y^{i-1}}(dx_i|x^{i-1},y^{i-1})\nonumber
\end{align}
\noi {\bf Sequential and Non-Anticipative Rate Distortion Function.} Based on $a)$ or $b)$ the operational definition of sequential and non-anticipative rate distortion function is given by expression
\begin{align}
R^c(D)\tri\lim_{n\rightarrow\infty}\inf_{\overrightarrow{P}_{Y^n|X^n}(\cdot|\cdot)\in\overrightarrow{Q}(D)}\frac{1}{n+1}I({X^n}\rightarrow{Y^n})\label{equation1e}
\end{align}
 where $\overrightarrow{Q}(D)$ is the distortion fidelity constraint and 
 \begin{align*}
&\overrightarrow{P}_{Y^n|X^n}(dy^n|x^n)\tri\otimes_{i=0}^n{P}_{Y_i|Y^{i-1},X^i}(dy_i|y^{i-1},x^i),\\
&{P}_{X_i|X^{i-1},Y^{i-1}}(dx_i|x^{i-1},y^{i-1})=P_{X_i|X^{i-1}}(dx_i|x^{i-1})-a.s.
 \end{align*}
The complete investigation of existence, characterization, and properties of the above extremum problems requires extensive analysis of the functional $\mathbb{I}_{X^n\rightarrow{Y^n}}(\cdot,\cdot)$ as defined in (\ref{equation1c}). This is analogous to capacity of channels without feedback which involves maximization of mutual information $I({X^n};{Y^n})$ over the power constraint set, and to classical rate distortion function which involves minimization of mutual information $I({X^n};{Y^n})$ over the fidelity constraint.
However, mutual information $I(X^n;Y^n)\equiv\mathbb{I}_{X^n;Y^n}(P_{X^n},P_{Y^n|X^n})$,  inherits from its information divergence definition $I(X^n;Y^n)\tri\mathbb{D}(P_{X^n,Y^n}||P_{X^n}\times{P}_{Y^n})$, several important functional properties such as convexity, concavity, lower semicontinuity, etc.
These properties are vital both for finite alphabet spaces, as well as abstract alphabet spaces \cite{csiszar92,csiszar74}. The difficulty associated with directed information $I(X^n\rightarrow{Y^n})$, rises from the fact that this information measure (\ref{equation1a})-(\ref{equation1c}) is a functional $\mathbb{I}_{X^n\rightarrow{Y}^n}(\cdot,\cdot)$ of the collection of conditional distributions $\{P_{X_i|X^{i-1},Y^{i-1}}(\cdot|\cdot,\cdot):i=0,1,\ldots,n\}$ and $\{P_{Y_i|Y^{i-1},X^i}(\cdot|\cdot,\cdot):i=0,1,\ldots,n\}$.\\
The objective of this paper is to address the following questions, when ${\cal X}_{0,n}$ and ${\cal Y}_{0,n}$ are complete separable metric spaces (Polish spaces).
\begin{itemize}
\item[1.] Is there an equivalent directed information definition expressed via information  divergence $\mathbb{D}(\cdot||\cdot)$ as a functional of two appropriate conditional distributions ${\bf P}(\cdot|{\bf y})$  on ${\cal X}^{\mathbb{N}}\tri\times_{i=0}^{\infty}{\cal X}_i$ for ${\bf y}=(y_0,y_1,\ldots)\in{\cal Y}^{\mathbb{N}}\tri\times_{i=0}^{\infty}{\cal Y}_i$ and ${\bf Q}(\cdot|{\bf x})$ on ${\cal Y}^{\mathbb{N}}$ for ${\bf x}\in{\cal X}^{\mathbb{N}}$ which uniquely define $\{P_{X_i|X^{i-1},Y^{i-1}}:i=0,1,\ldots\}$ and $\{P_{Y_i|Y^{i-1},X^i}:i=0,1,\ldots\}$, respectively, and vice-versa?
\item[2.] Is directed information convex and concave functional with respect to the conditional distributions ${\bf P}(\cdot|{\bf y})$ and ${\bf Q}(\cdot|{\bf x})$?
\item[3.] Is directed information a lower semicontinuous functional of the conditional distributions ${\bf P}(\cdot|{\bf y})$ and ${\bf Q}(\cdot|{\bf x})$?
\item[4.] What are appropriate conditions for existence of the maximizing encoder admissible distributions and minimizing distortion admissible distributions?
\end{itemize}
This paper answers the above questions by invoking the topology of weak convergence of probability measures on Polish spaces and Prohorov's theorems.\\
The paper is organized as follows. Section~\ref{causal_channels} provides the construction of two equivalent definitions of causal channels on abstract spaces while in Section~\ref{directed_information} the main properties of directed information are given. Finally, in Section~\ref{extremum_problems} the extremum problems (\ref{equation1d}), (\ref{equation1e}) are discussed. The derivations of theorems are outlined since they are quite lengthy.
\section{Causal Channels on Abstract Spaces}\label{causal_channels}
In this section, the aim is to establish two equivalent definitions of conditional distributions or basic processes, which define any probabilistic channel with causal feedback, that relates causally the input-output behavior of any channel. This formulation is necessary to investigate questions 1.--4.\\
Let $\mathbb{N} \tri \{0,1,2,\ldots\},$ and $\mathbb{N}^n \tri \{0,1,2,\ldots,n\}.$ Introduce two sequence of spaces $\{({\cal X}_n,{\cal B}({\cal X }_n)):n\in\mathbb{N}\}$ and $\{({\cal Y}_n,{\cal B}({\cal Y}_n)):n\in\mathbb{N}\},$ where ${\cal X}_n,{\cal Y}_n, n\in\mathbb{N}$ are topological spaces, and ${\cal B}({\cal X}_n)$ and ${\cal B}({\cal Y}_n)$ are Borel $\sigma-$algebras of subsets of ${\cal X}_n$ and ${\cal Y}_n,$ respectively. 
Points in ${\cal X}^{\mathbb{N}}\tri{{\times}_{n\in\mathbb{N}}}{\cal X}_n,$ ${\cal Y}^{\mathbb{N}}\tri{\times_{n\in\mathbb{N}}}{\cal Y}_n$ are denoted by ${\bf x}\tri\{x_0,x_1,\ldots\}\in{\cal X}^{\mathbb{N}},$ ${\bf y}\tri\{y_0,y_1,\ldots\}\in{\cal Y}^{\mathbb{N}},$ respectively, while their restrictions to finite coordinates by $x^n\tri\{x_0,x_1,\ldots,x_n\}\in{\cal X}_{0,n},$ $y^n\tri\{y_0,y_1,\ldots,y_n\}\in{\cal Y}_{0,n},$ for $n\in\mathbb{N}$.\\
Let ${\cal B}({\cal X}^{\mathbb{N}})\tri\odot_{i\in\mathbb{N}}{\cal B}({\cal X}_i)$ denote the $\sigma-$algebra on ${\cal X}^{\mathbb{N}}$ generated by cylinder sets $\{{\bf x}=(x_0,x_1,\ldots)\in{\cal X}^{\mathbb{N}}:x_0\in{A}_0,x_1\in{A}_1,\ldots,x_n\in{A}_n\}, A_i\in{\cal B}({\cal X}_i), 0\leq{i}\leq{n}, n\geq1$, and similarly for ${\cal B}({\cal Y}^{\mathbb{N}})\tri\odot_{i\in\mathbb{N}}{\cal B}({\cal Y}_i)$. Hence, ${\cal B}({\cal X}_{0,n})$ and ${\cal B}({\cal Y}_{0,n})$ denote the $\sigma-$algebras of cylinder sets in ${\cal X}^{\mathbb{N}}$ and ${\cal Y}^{\mathbb{N}},$ respectively, with bases over $A_i\in{\cal B}({\cal X}_i)$, and $B_i\in{\cal B}({\cal Y}_i),~0\leq{i}\leq{n}$, respectively.\\
\noi{\bf Backward or Feedback Channel.} Suppose for each $n\in\mathbb{N},$ the distributions $\{p_n(dx_n|x^{n-1},y^{n-1}):n\in\mathbb{N}\}$ with $p_0(dx_0|x^{-1},y^{-1})\tri{p}_0(x_0)$ satisfy the following conditions.\\
{\bf i)} For $n\in\mathbb{N},$ $p_n(\cdot|x^{n-1},y^{n-1})$ is a probability measure on ${\cal B}({\cal X }_n);$\\
{\bf ii)} For $n\in\mathbb{N},$ $A_n\in{\cal B}({\cal X }_n)$, $p_n(A_n|x^{n-1},y^{n-1})$ is $\odot^{n-1}_{i=0}{\cal B}({\cal X }_i)\odot{\cal B}({\cal Y}_i)-$measurable in $x^{n-1}\in{\cal X}_{0,n-1},$ $y^{n-1}\in{\cal Y}_{0,n-1}.$\\
Given the collection $\{p_n(dx_n|x^{n-1},y^{n-1}):n\in\mathbb{N}\}$ satisfying conditions {\bf i)}, {\bf ii)}, one can construct a family of distributions on  $({\cal X}^{\mathbb{N}},{\cal B}({\cal X}^{\mathbb{N}}))\tri\big(\times_{i\in\mathbb{N}}{\cal X}_i,\odot_{i\in\mathbb{N}}{\cal B}({\cal X }_i)\big)$ as follows.\\
Let $C\in{\cal B}({\cal X}_{0,n})$ be a cylinder set of the form $
C\tri\big\{{\bf x}\in{\cal X}^{\mathbb{N}}:x_0\in{C_0},x_1\in{C_1},\ldots,x_n\in{C_n}\big\},~C_i\in{\cal B}({\cal X }_i),~0\leq i \leq n$.
Define a family of measures ${\bf P}(\cdot|{\bf y})$ on ${\cal B}({\cal X}^{\mathbb{N}})$ by
\begin{align}
{\bf P}(C|{\bf y})&\tri\int_{C_0}p_0(dx_0)\ldots\int_{C_n}p_n(dx_n|x^{n-1},y^{n-1})\label{equation2}\\
&\equiv{\overleftarrow{P}}_{0,n}(C_{0,n}|y^{n-1}),~C_{0,n}=\times_{i=0}^n{C_i}\label{equation4a}
\end{align}
The notation ${\overleftarrow{P}}_{0,n}(\cdot|y^{n-1})$ is used to denote the restriction of the measure ${\bf P}(\cdot|{\bf y})$ on cylinder sets $C\in{\cal B}({\cal X}_{0,n})$, for $n\in\mathbb{N}$.\\
Thus, if conditions {\bf i)} and {\bf ii)} hold then for each ${\bf y}\in{\cal Y}^{\mathbb{N}},$ the right hand side of $(\ref{equation2})$ defines a consistent family of finite-dimensional distribution on $({\cal X}^{\mathbb{N}},{\cal B}({\cal X}^{\mathbb{N}}))$, and hence there exists a unique measure on $({\cal X}^{\mathbb{N}},{\cal B}({\cal X}^{\mathbb{N}})),$ from which $p_n(dx_n|x^{n-1},y^{n-1})$ is obtained. This leads to the first, usual definition of a feedback channel, as a family of functions $p_n(dx_n|x^{n-1},y^{n-1})$ satisfying conditions ${\bf i)}$ and ${\bf ii)}.$\\
An alternative, equivalent definition of a feedback channel is established as follows. Introduce the assumption\\ ${\bf iii)}$ $\{{\cal X}_n:n\in\mathbb{N}\}$ are complete separable metric spaces (Polish Spaces) and $\{{\cal B}({\cal X}_n):n\in\mathbb{N}\}$ are the $\sigma-$algebras of Borel sets.\\
Consider a family of measures ${\bf P}(\cdot|{\bf y})$ on $({\cal X}^{\mathbb{N}},{\cal B}({\cal X}^{\mathbb{N}}))$ satisfying the following consistency condition.\\
{\bf C1}:~~If $E\in{\cal B}({\cal X}_{0,n})$, then ${\bf P}(E|{\bf y})$ is ${\cal B}({\cal Y}_{0,n-1})-$measurable function of ${\bf y}\in{\cal Y}^{\mathbb{N}}$.\\
Then, by assumption ${\bf iii)}$, for any family of measures ${\bf P}(\cdot|{\bf y})$ satisfying {\bf C1} one can construct a collection of versions of conditional distributions  $\{p_n(dx_n|x^{n-1},y^{n-1}):n\in\mathbb{N}\}$ satisfying conditions ${\bf i)}$ and ${\bf ii)}$ which are connected with ${\bf P}(\cdot|{\bf y})$ via relation $(\ref{equation2}).$\\
Therefore, for Polish Spaces  $\{{\cal X}_n:n\in\mathbb{N}\}$ the second equivalent definition is given by a family of measures ${\bf P}(\cdot|{\bf y})$ on $({\cal X}^{\mathbb{N}},{\cal B}({\cal X}^{\mathbb{N}}))$ depending parametrically on ${\bf y}\in{\cal Y}^{\mathbb{N}}$ and satisfying the consistency condition {\bf C1}.\\
The point to be made here is that the second equivalent definition of a feedback channel, together with similar definition for the forward channel is convenient to define directed information via relative entropy, similar to the mutual information definition, and extend well-known functional properties of mutual information to directed information.\\
\noi{\bf Forward Channel.} The previous methodology is repeated for the collection of functions $\{q_n(dy_n|y^{n-1},x^n):n\in\mathbb{N}\}$ which satisfy the following conditions.\\
{\bf iv)} For $n\in\mathbb{N},$ $q_n(\cdot|y^{n-1},x^{n})$ is a probability measure on ${\cal B}({\cal Y}_n);$\\
{\bf v)} For $n\in\mathbb{N}$, $B_n\in{\cal B}({\cal Y}_n)$, $q_n(B_n|y^{n-1},x^{n})$ is $\odot^{n-1}_{i=0}{\cal B}({\cal Y}_i)\odot_{i=0}^n{\cal B}({\cal X}_i)-$measurable function of $x^{n}\in{\cal X}_{0,n},$ $y^{n-1}\in{\cal Y}_{0,n-1}$.\\
Similarly as before, given a cylinder set $D\tri\Big\{{\bf y}\in{\cal Y}^{\mathbb{N}}:y_0{\in}D_0,y_1{\in}D_1,\ldots,y_n{\in}D_n\Big\},~D_i\in{\cal B}({\cal Y}_i),~0\leq i \leq n$, define a family of measures on ${\cal B}({\cal Y}^{\mathbb{N}})$ by
\begin{align}
{\bf Q}(D|{\bf x})&\tri\int_{D_0}q_0(dy_0|x_0)\ldots\int_{D_n}q_n(dy_n|y^{n-1},x^n)\label{equation4}\\
&\equiv{\overrightarrow{Q}}_{0,n}(D_{0,n}|x^n),~D_{0,n}=\times_{i=0}^n{D_i}\label{equation4b}
\end{align}
Similarly as before, there exists a unique measure on $({\cal Y}^{\mathbb{N}},{\cal B}({\cal Y}^{\mathbb{N}}))$ for which the family of distributions $\{q_n(dy_n|y^{n-1},x^n):n\in\mathbb{N}\}$ is obtained. Introduced the assumption\\
${\bf vi)}$ $\{{\cal Y}_n:n\in\mathbb{N}\}$ are Polish Spaces and $\{{\cal B}({\cal Y}_n):n\in\mathbb{N}\}$ are the $\sigma-$algebras of Borel sets.\\
Consider a family of measures ${\bf Q}(D|{\bf x})$ satisfying the following consistency condition.\\
{\bf C2}: If $F\in{\cal B}({\cal Y}_{0,n}),$ then ${\bf Q}(F|{\bf x})$ is ${\cal B}({\cal X}_{0,n})-$measurable function of ${\bf x}\in{\cal X}^{\mathbb{N}}.$\\
Then, by assumption ${\bf vi)}$, for any family of measures ${\bf Q}(\cdot|{\bf x})$ on $({\cal Y}^{\mathbb{N}},{\cal B}({\cal Y}^{\mathbb{N}}))$ satisfying consistency condition {\bf C2} one can construct a collection of functions $\{q_n(dy_n|y^{n-1},x^n):n\in\mathbb{N}\}$ satisfying conditions ${\bf iv)}$ and ${\bf v)}$ which are connected with ${\bf Q}(\cdot|{\bf x})$ via relation $(\ref{equation4})$. Note that Kolmogorov's extension theorem guarantees the construction of countable additive probability measures for both ${\bf P}(\cdot|{\bf y})$ and ${\bf Q}(\cdot|{\bf x})$.\\
Given the basic measures ${\bf P}(\cdot|{\bf y})$ on ${\cal X}^{\mathbb{N}}$ and ${\bf Q}(\cdot|{\bf x})$ on ${\cal Y}^{\mathbb{N}}$ satisfying consistency condition ${\bf C1}$ and ${\bf C2}$, respectively, construct the collections of conditional distributions as follows.\\
Let $A^{(n)}=\{{\bf x}:x_n{\in}A\},$ $A\in{\cal B}({\cal X}_n)$ and $B^{(n)}=\{{\bf y}:y_n{\in}B\},$ $B\in{\cal B}({\cal Y}_n).$ In addition, let ${\bf P}(A^{(n)}|{\bf y}|{{\cal B}({\cal X}_{0,n-1})})$ denote the conditional probability of $A^{(n)}$ with respect to ${\cal B}({\cal X}_{0,n-1})$ calculated on the probability space $\big({\cal X}^{\mathbb{N}},{\cal B}({\cal X}^{\mathbb{N}}),{\bf P}(\cdot|{\bf y})\big),$ and similarly for ${\bf Q}(B^{(n)}|{\bf x}|{{\cal B}({\cal Y}_{0,n-1})})$. Then
\begin{align*}
&\mathbb{P}\big\{X_n{\in}A|X^{n-1}=x^{n-1},Y^{n-1}=y^{n-1}\big\}\\
&={\bf P}\big(\{{\bf x}:x_n{\in}A\}|{\bf y}|{{\cal B}({\cal X}_{0,n-1})}\big)=p_n(A_n;x^{n-1},y^{n-1})-a.s.\\
&\mathbb{P}\big\{Y_n{\in}B|Y^{n-1}=y^{n-1},X^n=x^n\big\} \\
&={\bf Q}\big(\{{\bf y}:y_n{\in}B\}|{\bf x}|{{\cal B}({\cal Y}_{0,n-1})}\big)=q_n(B_n;y^{n-1},x^n)-a.s.
\end{align*}
Note that $p_n(\cdot;\cdot,\cdot)\in{\cal Q}({\cal X}_n;{\cal X}_{0,n-1}\times{\cal Y}_{0,n-1})$ and $q_n(\cdot;\cdot,\cdot)\in{\cal Q}({\cal Y}_n;{\cal Y}_{0,n-1}\times{\cal X}_{0,n})$ are stochastic kernels \cite{dupuis-ellis97}, determined from ${\bf P}(\cdot|\cdot)$ and ${\bf Q}(\cdot|\cdot),$ respectively, (e.g., related via $(\ref{equation2})$, $(\ref{equation4})$).\\
The distribution of RV's $\{(X_i,Y_i):i\in\mathbb{N}\}$ is defined by
\begin{align}
{P}&\big\{X_0{\in}A_0,Y_0\in{B}_0,\ldots,X_n{\in}A_n,Y_n{\in}B_n\big\}\nonumber\\
&\tri\int_{A_0}p_0(dx_0)\int_{B_0}q_0(dy_0;x_0)\ldots\int_{B_n}q_n(dy_n;y^{n-1},x^n)\nonumber
\end{align}
Hence, for any ${\bf P}(\cdot|\cdot)$ and ${\bf Q}(\cdot|\cdot)$ satisfying consistency conditions there exist a probability space and a sequence of RV's $\{(X_i,Y_i):i\in\mathbb{N}\}$ defined on it, whose joint probability distribution is defined uniquely via ${\bf P}(\cdot|\cdot)$ and ${\bf Q}(\cdot|\cdot)$.
\section{Directed Information Properties and Compactness}\label{directed_information}
\par In this section, directed information $I(X^n\rightarrow{Y^n})$ will be defined via relative entropy, using the basic measures ${\bf P}(\cdot|{\bf y})$  and ${\bf Q}(\cdot|{\bf x})$, and identify its properties.
Define
\begin{align}
&{\cal Q}^{\bf C1}({\cal X}^{\mathbb{N}};{\cal Y}^{\mathbb{N}})\tri\Big\{
{\bf P}(\cdot|{\bf y})\in{\cal M}_1({\cal X}^{\mathbb{N}}):{\bf P}(\cdot|{\bf y})~\mbox{are regular}\nonumber\\
&\mbox{ probability measures and satisfy consistency condition {\bf C1}}\Big\}.\nonumber\\
&{\cal Q}^{\bf C2}({\cal Y}^{\mathbb{N}};{\cal X}^{\mathbb{N}})\tri\Big\{{\bf Q}(\cdot|{\bf x})\in{\cal M}_1({\cal Y}^{\mathbb{N}}):{\bf Q}(\cdot|{\bf x})~\mbox{ are regular} \nonumber\\
&\mbox{probability measures and satisfy consistency condition {\bf C2}}\Big\}.\nonumber
\end{align}
Given conditional distributions ${\bf P}(\cdot|\cdot)\in{\cal Q}^{\bf C1}({\cal X}^{\mathbb{N}};{\cal Y}^{\mathbb{N}})$ and ${\bf Q}(\cdot|\cdot)\in{\cal Q}^{\bf C2}({\cal Y}^{\mathbb{N}};{\cal X}^{\mathbb{N}})$ define the following measures.\\
{\bf P1}: The joint distribution on ${\cal X}^{\mathbb{N}}\times{\cal Y}^{\mathbb{N}}$ defined uniquely by
\begin{align}
({\overleftarrow P}_{0,n}&\otimes{\overrightarrow Q}_{0,n})(\times^n_{i=0}A_i{\times}B_i),A_i\in{\cal B}({\cal X}_i),~B_i\in{\cal B}({\cal Y}_i), \nonumber  \\
&{\tri}\mathbb{P}\big\{X_0{\in}A_0,Y_0\in{B}_0,\ldots,X_n{\in}A_n,Y_n{\in}B_n\big\}\label{equation18}
\end{align}
{\bf P2}: The marginal distributions on ${\cal X}^{\mathbb{N}}$ defined uniquely by
\begin{align}
&\mu_{0,n}(\times^n_{i=0}A_i),~A_i\in{\cal B}({\cal X}_i), ~ 1\leq i \leq n\nonumber\\
&\tri\mathbb{P}\{X_0\in{A}_0, Y_0\in{\cal Y}_0,\ldots, X_n\in{A}_n, Y_n\in{\cal Y}_n\},\nonumber\\
&=({\overleftarrow P}_{0,n}\otimes{\overrightarrow Q}_{0,n})(\times^n_{i=0}(A_i\times{\cal Y}_i))\nonumber
\end{align}
{\bf P3}: The marginal distributions on ${\cal Y}^{\mathbb{N}}$ defined uniquely for $B_i\in{\cal B}({\cal Y}_i),~1\leq i \leq n$ by
\begin{align}
\nu_{0,n}(\times^n_{i=0}B_i)=({\overleftarrow P}_{0,n}\otimes{\overrightarrow Q}_{0,n})(\times^n_{i=0}({\cal X}_i\times{B}_i))\nonumber
\end{align}
{\bf P4}: The measure ${\overrightarrow\Pi}_{0,n}:{\cal B}({\cal X}_{0,n})\odot{\cal B}({\cal Y}_{0,n})\mapsto[0,1]$ defined uniquely for $A_i\in{\cal B}({\cal X}_i)$,~$B_i\in{\cal B}({\cal Y}_i)$,~$1\leq i \leq n$ by
\begin{align}
{\overrightarrow\Pi}_{0,n}(\times^n_{i=0}(A_i{\times}B_i))&\tri({\overleftarrow P}_{0,n}\otimes\nu_{0,n})(\times^n_{i=0}(A_i{\times}B_i))\nonumber
\end{align}
{\bf P5}: The measure ${\overleftarrow\Pi}_{0,n}:{\cal B}({\cal Y}_{0,n})\odot{\cal B}({\cal X}_{0,n})\mapsto[0,1]$ defined uniquely for~$A_i\in{\cal B}({\cal X}_i)$,~$B_i\in{\cal B}({\cal Y}_i)$,~$1\leq i\leq n$  by
\begin{align}
{\overleftarrow\Pi}_{0,n}(\times^n_{i=0}(A_i{\times}B_i))&\tri(\mu_{0,n}\otimes{\overrightarrow Q}_{0,n})(\times^n_{i=0}(A_i{\times}B_i))\nonumber
\end{align}
\subsection{Directed Information}
Let ${\bf P}(\cdot|\cdot)\in{\cal Q}^{\bf C1}({\cal X}^{\mathbb{N}};{\cal Y}^{\mathbb{N}})$ and ${\bf Q}(\cdot|\cdot)\in{\cal Q}^{\bf C2}({\cal X}^{\mathbb{N}};{\cal Y}^{\mathbb{N}}).$ By invoking the definition of directed information (\ref{equation1a}) or (\ref{equation1b}), it can be shown that
\begin{align}
&I(X^n\rightarrow{Y}^n) = \mathbb{D}({\overleftarrow P}_{0,n} \otimes {\overrightarrow Q}_{0,n}||{\overrightarrow\Pi}_{0,n})\label{equation33}\\
&=\int\log \Big( \frac{{\overrightarrow Q}_{0,n}(d y^n|x^n)}{\nu_{0,n}(dy^n)}\Big)({\overleftarrow P}_{0,n}\otimes {\overrightarrow Q}_{0,n})(dx^n,dy^n) \nonumber   \\
&\equiv{\mathbb{I}}_{X^n\rightarrow{Y^n}}({\overleftarrow P}_{0,n}, {\overrightarrow Q}_{0,n}) \label{equation7a}       
\end{align}
The right hand side of (\ref{equation33}) follows from repeated application of chain rule of relative entropy \cite{dupuis-ellis97}, while (\ref{equation7a}) follows from the fact that ${\overleftarrow P}_{0,n} \otimes {\overrightarrow Q}_{0,n} << {\overleftarrow P}_{0,n} \otimes \nu_{0,n}$ if and only if ${\overrightarrow Q}_{0,n}(\cdot|x^n) << \nu_{0,n}(\cdot)$ for $\mu_{0,n}-$almost all $x^n \in {\cal X}_{0,n}$. Further, if  ${\overleftarrow P}_{0,n} \otimes {\overrightarrow Q}_{0,n}  << {\overleftarrow P}_{0,n} \otimes \nu_{0,n}$ then the Radon-Nikodym derivative $ \frac {({\overleftarrow P}_{0,n} \otimes {\overrightarrow Q}_{0,n})}{ ({\overleftarrow P}_{0,n}\otimes \nu_{0,n} )}(x^n,y^n)$ represents a version of $\frac {{\overrightarrow Q}_{0,n}(\cdot|x^n) }{  \nu_{0,n} (\cdot) }(y^n)$, $\mu_{0,n}-a.s$ for all  $x^n \in {\cal X}_{0,n}$.\\
The notation ${\mathbb{I}}_{X^n\rightarrow{Y^n}}(\cdot,\cdot)$ indicates the functional dependence of $I(X^n\rightarrow{Y^n})$ on $\{{\overleftarrow P}_{0,n}, {\overrightarrow Q}_{0,n}\}$.
\subsection{Convexity and Concavity of Directed Information}
\noi Let ${\cal Q}^{\bf C1}({\cal X}_{0,n};{\cal Y}_{0,n-1})$, ${\cal Q}^{\bf C2}({\cal Y}_{0,n};{\cal X}_{0,n})$   be the restrictions of ${\cal Q}^{\bf C1}({\cal X}^{\mathbb{N}};{\cal Y}^{\mathbb{N}})$ and ${\cal Q}^{\bf C2}({\cal Y}^{\mathbb{N}};{\cal X}^{\mathbb{N}})$, respectively, to cylinder sets with bases over $A_i\in{\cal B}({\cal X}_i)$,  and $B_i\in{\cal B}({\cal Y}_i)$, $i=0,1,\ldots,n$. These are  regular conditional distributions.
\begin{theorem}\label{convexity1}
Let $\{({\cal X}_n,{\cal B}({\cal X}_n)):n\in{\mathbb{N}}\},$ $\{({\cal Y}_n,{\cal B}({\cal Y}_n)):n\in{\mathbb{N}}\}$ be Polish spaces. Then
\begin{itemize}
\item[1)] ${\cal Q}^{\bf C1}({\cal X}_{0,n};{\cal Y}_{0,n-1})$, ${\cal Q}^{\bf C2}({\cal Y}_{0,n};{\cal X}_{0,n})$ are convex sets.
\item[2)] ${\mathbb{I}}_{X^n\rightarrow{Y}^n}({\overleftarrow P}_{0,n},{\overrightarrow Q}_{0,n})$ is a convex functional of ${\overrightarrow Q}_{0,n}\in{\cal Q}^{\bf C2}({\cal Y}_{0,n};{\cal X}_{0,n})$ for a fixed ${\overleftarrow P}_{0,n}\in{\cal Q}^{\bf C1}({\cal X}_{0,n};{\cal Y}_{0,n-1})$.
\item[3)] ${\mathbb{I}}_{X^n\rightarrow{Y}^n}({\overleftarrow P}_{0,n},{\overrightarrow Q}_{0,n})$ is a concave functional of ${\overleftarrow P}_{0,n}\in{\cal Q}^{\bf C1}({\cal X}_{0,n};{\cal Y}_{0,n-1})$ for a fixed ${\overrightarrow Q}_{0,n}\in{\cal Q}^{\bf C2}({\cal Y}_{0,n};{\cal X}_{0,n}).$
\end{itemize}
\end{theorem}
{\IEEEproof} 1) Utilize the convexity of regular conditional distributions, and then the consistency condition ${\bf C1}$, ${\bf C2}$. 2), 3), follow from  log-sum formulae.
{\endIEEEproof}
\subsection{Lower semicontinuity-Continuity of Directed Information}
\par This part discusses the lower-semicontinuity and continuity of directed information as a functional of ${\overleftarrow P}_{0,n}(\cdot|y^{n-1})\in{\cal Q}^{\bf C1}({\cal X}_{0,n};{\cal Y}_{0,n-1})$ and $\overrightarrow Q_{0,n}(\cdot|x^n)\in{\cal Q}^{\bf C2}({\cal Y}_{0,n};{\cal X}_{0,n})$. Before establishing the main results, sufficient conditions for weak compactness of the set of measures ${{\cal Q}^{\bf C1}({\cal X}_{0,n};{\cal Y}_{0,n-1})}$, ${{\cal Q}}^{\bf C2}({\cal Y}_{0,n};{\cal X}_{0,n}),$ and joint and marginal measures are given.
\begin{theorem}\label{weak_convergence}
{\bf Part A.} Let ${\cal Y}_{0,n}$ be a compact Polish space and ${\cal X}_{0,n}$ a Polish space. Assume ${\overleftarrow P}_{0,n}(\cdot|y^{n-1})\in{\cal Q}^{\bf C1}({\cal X}_{0,n};{\cal Y}_{0,n-1})$ satisfy the following condition.\\
{\bf CA}: For all $g(\cdot){\in}BC({\cal X}_{0,n})$, where $BC({\cal X}_{0,n})$ denotes the set of bounded continuous real-valued functions on ${\cal X}_{0,n}$,
\begin{align}
(x^{n-1},y^{n-1})\longmapsto\int_{{\cal X}_n}g(x)p_n(dx;x^{n-1},y^{n-1})\in\mathbb{R}\label{condition1}
\end{align}
is jointly continuous in $(x^{n-1},y^{n-1})\in{\cal X}_{0,n-1}\times{\cal Y}_{0,n-1}$.\\
Then the following weak convergence results hold.
\begin{itemize}
\item[A1)] Let ${\overleftarrow P}_{0,n}(\cdot|y^{n-1})\in{\cal Q}^{\bf C1}({\cal X}_{0,n};{\cal Y}_{0,n-1})$ and $\big\{{\overrightarrow Q}_{0,n}^{\alpha}(\cdot|{ x^n})\big\}_{\alpha \geq 1}\in {\cal Q}^{\bf C2}({\cal Y}_{0,n};{\cal X}_{0,n}).$ Then the joint measure $({\overleftarrow P}_{0,n}\otimes{\overrightarrow Q}_{0,n}^{\alpha})(dx^n,dy^n)\buildrel w \over\Longrightarrow({\overleftarrow P}_{0,n}\otimes{\overrightarrow Q}_{0,n}^0)(dx^n,dy^n),$ where ${\overrightarrow Q}_{0,n}^0(\cdot|x^n)\in{\cal Q}^{\bf C2}({\cal Y}_{0,n};{\cal X}_{0,n}).$
\item[A2)] Let ${\overleftarrow P}_{0,n}(\cdot|y^{n-1})\in{\cal Q}^{\bf C1}({\cal X}_{0,n};{\cal Y}_{0,n-1})$ and $\big\{{\overrightarrow Q}_{0,n}^{\alpha}(\cdot|{ x^n})\big\}_{\alpha \geq 1}\in{\cal Q}^{\bf C2}({\cal Y}_{0,n};{\cal X}_{0,n})$ and define the family of joint measures $\big\{({\overleftarrow P}_{0,n}\otimes{\overrightarrow Q}_{0,n}^\alpha)(dx^n,dy^n)\big\}_{\alpha \geq 1}$ having marginals $\{\nu_{0,n}^{\alpha}\}_{\alpha \geq 1}$ on ${\cal Y}_{0,n}$ and $\{\mu_{0,n}^{\alpha}\}_{\alpha \geq 1}$ on ${\cal X}_{0,n}$. Then $\nu_{0,n}^{\alpha}(dy^n)\buildrel w \over\Longrightarrow\nu_{0,n}^0(dy^n)$ and $\mu_{0,n}^{\alpha}(dx^n)\buildrel w \over\Longrightarrow\mu_{0,n}^0(dx^n)$ where $\nu_{0,n}^0\in{\cal M}_1({\cal Y}_{0,n})$ and $\mu_{0,n}^0\in{\cal M}_1({\cal 	 X}_{0,n})$ are the marginals of $({\overleftarrow P}_{0,n}\otimes{\overrightarrow Q}_{0,n}^0)(dx^n,dy^n).$
\item[A3)] The sets of measures ${\cal Q}^{\bf C1}({\cal X}_{0,n};{\cal Y}_{0,n-1})$, and ${\cal Q}^{\bf C2}({\cal Y}_{0,n};{\cal X}_{0,n})$ are weakly compact.
\item[A4)] Let ${\overleftarrow P}_{0,n}(\cdot|y^{n-1})\in{\cal Q}^{\bf C1}({\cal X}_{0,n};{\cal Y}_{0,n-1}),$ $\big\{{\overrightarrow Q}_{0,n}^{\alpha}(\cdot|{ x^n})\big\}_{\alpha \geq 1}\in{\cal Q}^{\bf C2}({\cal Y}_{0,n};{\cal X}_{0,n}),$ and $\{\nu_{0,n}^{\alpha}\}_{\alpha \geq 1}$ the marginals of $\big\{({\overleftarrow P}_{0,n}\otimes{\overrightarrow Q}_{0,n}^\alpha)(dx^n,dy^n)\big\}_{\alpha \geq 1}.$ Then ${\overrightarrow\Pi}_{0,n}^\alpha(dx^n,dy^n)\equiv{\overleftarrow P}_{0,n}(dx^n|dy^{n-1})\otimes\nu_{0,n}^{\alpha}(dy^n)\buildrel w \over\Longrightarrow{\overleftarrow P}_{0,n}(dx^n|dy^{n-1})\otimes\nu_{0,n}^0(dy^n)\equiv{\overrightarrow\Pi}_{0,n}^0(dx^n,dy^n),$ where $\nu_{0,n}^0\in{\cal M}_1({\cal Y}_{0,n})$ is the weak limit of $\nu_{0,n}^{\alpha}\in{\cal M}_1({\cal Y}_{0,n})$.
\end{itemize}
{\bf Part B.} Let ${\cal X}_{0,n}$ be a compact Polish space and ${\cal Y}_{0,n}$ a Polish space. Assume ${\overrightarrow Q}_{0,n}(\cdot|x^{n})\in{\cal Q}^{\bf C2}({\cal Y}_{0,n};{\cal X}_{0,n})$ satisfy the following condition.\\
{\bf CB}: For all $h(\cdot){\in}BC({\cal Y}_{0,n})$, the function
\begin{align}
(x^{n},y^{n-1})\longmapsto\int_{{\cal Y}_n}h(y)q_n(dy;y^{n-1},x^n)\in\mathbb{R}\label{condition2}
\end{align}
is jointly continuous in $(x^{n},y^{n-1})\in{\cal X}_{0,n}\times{\cal Y}_{0,n-1}$.\\
The statements of {\bf Part A} hold by interchanging ${\overleftarrow Q}_{0,n}$ with ${\overleftarrow P}_{0,n}$, $\nu_{0,n}$ with $\mu_{0,n}$, ${\overrightarrow\Pi}_{0,n}$ with ${\overleftarrow\Pi}_{0,n}$.
\end{theorem}
{\IEEEproof} The proof is quite lengthy and it is based on Prohorov's theorem relating tightness and weak compactness of a family of probability measures \cite{dupuis-ellis97}.{\endIEEEproof}
The results of Theorem~$\ref{weak_convergence}$ are sufficient to establish lower semicontinuity of directed information $I(X^n\rightarrow{Y}^n)\equiv{\mathbb{I}}_{X^n\rightarrow{Y}^n}({\overleftarrow P}_{0,n},{\overrightarrow Q}_{0,n})$.
\begin{theorem}\label{lower_semicontinuity}
1) Suppose the conditions in Theorem~\ref{weak_convergence}, {\bf Part A} hold.
Then
${\mathbb{I}}_{X^n\rightarrow{Y^n}}({\overleftarrow P}_{0,n}, {\overrightarrow Q}_{0,n})$ is lower semicontinuous on ${\overrightarrow Q}_{0,n}\in{\cal Q}^{\bf C2}({\cal Y}_{0,n};{\cal X}_{0,n})$ for fixed ${\overleftarrow P}_{0,n}\in{\cal Q}^{\bf C1}({\cal X}_{0,n};{\cal Y}_{0,n-1})$.\\
2) Suppose the conditions in Theorem~\ref{weak_convergence}, {\bf Part B} hold.
Then ${\mathbb{I}}_{X^n\rightarrow{Y^n}}({\overleftarrow P}_{0,n}, {\overrightarrow Q}_{0,n})$ is lower semicontinuous on ${\overleftarrow P}_{0,n}\in{\cal Q}^{\bf C1}({\cal X}_{0,n};{\cal Y}_{0,n-1})$ for fixed ${\overrightarrow Q}_{0,n}\in{\cal Q}^{\bf C2}({\cal Y}_{0,n};{\cal X}_{0,n})$.
\end{theorem}
{\IEEEproof} Utilizes (\ref{equation33}), Theorem~{\ref{weak_convergence}}, and lower semicontinuity of relative entropy.{\endIEEEproof}
For capacity problems, it is desirable to identify conditions so that ${\mathbb{I}}_{X^n\rightarrow{Y^n}}({\overleftarrow P}_{0,n}, {\overrightarrow Q}_{0,n})$ as a function of $\overleftarrow{P}_{0,n}$ for fixed $\overrightarrow{Q}_{0,n}$ is either upper semicontinuous or continuous. 
\begin{theorem}\label{upper_semicontinuity}
Consider a forward channel ${\overrightarrow Q}_{0,n}(\cdot|x^n)\in{\cal Q}^{\bf C2}({\cal Y}_{0,n};{\cal X}_{0,n}),$ and a closed family of feedback channels ${\cal Q}^{c,\bf C1}({\cal X}_{0,n};{\cal Y}_{0,n-1})\subseteq{\cal Q}^{\bf C1}({\cal X}_{0,n};{\cal Y}_{0,n-1}).$ Suppose there exists a family of measures $\bar{\nu}_{0,n}(dy^n)$ on $({\cal Y}_{0,n},{\cal B}({\cal Y}_{0,n}))$ such that ${\overrightarrow Q}_{0,n}(\cdot|x^n)\ll{\bar{\nu}}_{0,n}(dy^n)$ with Radon-Nikodym derivatives $\xi_{\bar{\nu}_{0,n}}(x^n,y^n)\tri\frac{{\overrightarrow Q}_{0,n}(\cdot|x^n)}{\bar{\nu}_{0,n}(\cdot)}(y^n),$ and\\
\noi 1) the family of Radon-Nikodym derivatives $\xi_{\bar{\nu}_{0,n}}(x^n,y^n)$ is continuous on ${\cal X}_{0,n}\times{\cal Y}_{0,n},$ and $\xi_{\bar{\nu}_{0,n}}(x^n,y^n)\log\xi_{\bar{\nu}_{0,n}}(x^n,y^n)$ is uniformly integrable over $\{\bar{\nu}_{0,n}\otimes{\overleftarrow P}_{0,n}:{\overleftarrow P}_{0,n}\in{\cal Q}^{c,\bf C1}({\cal X}_{0,n};{\cal Y}_{0,n-1})\}$;\\
\noi 2) for a fixed $y^n\in{\cal Y}_{0,n},$ the Radon-Nikodym derivative $\xi_{\bar{\nu}_{0,n}}(x^n,y^n)$ is uniformly integrable over ${\cal Q}^{c,\bf C1}({\cal X}_{0,n};{\cal Y}_{0,n-1}).$\\
\noi Then, the directed information ${\mathbb{I}}_{X^n\rightarrow{Y^n}}({\overleftarrow P}_{0,n},{\overrightarrow Q}_{0,n})$ as a functional of $\{{\overleftarrow P}_{0,n},{\overrightarrow Q}_{0,n}\}\in{\cal Q}^{c,\bf C1}({\cal X}_{0,n};{\cal Y}_{0,n-1})\times{\cal Q}^{\bf C2}({\cal Y}_{0,n};{\cal X}_{0,n})$ is bounded and weakly continuous over ${\cal Q}^{c,\bf C1}({\cal X}_{0,n};{\cal Y}_{0,n-1})$.
\end{theorem}
{\IEEEproof} Invoke (\ref{equation33}) and generalize related results in \cite{fozunbal}.
{\endIEEEproof}
\section{Extremum Problems of Directed Information}\label{extremum_problems}
\subsection{Existence of Capacity Achieving Distribution}
Consider a communication channel with memory and feedback ${\overrightarrow{Q}}_{0,n}(\cdot|x^n)\in{\cal Q}^{\bf C2}({\cal Y}_{0,n};{\cal X}_{0,n})$ and power constraint
\begin{align}
&\overleftarrow{\cal{P}}_{0,n}(P)\tri\big\{{\overleftarrow{P}}_{0,n}(\cdot|y^{n-1})\in{\cal Q}^{\bf C1}({\cal X}_{0,n};{\cal Y}_{0,n-1}):\nonumber\\
&\int{g}_{0,n}(x^n,y^{n-1})({\overleftarrow{Q}}_{0,n}\otimes{\overrightarrow{P}}_{0,n})(dx^n,dy^n)\leq{P}\big\}\nonumber
\end{align}
where for any $n\in\mathbb{N}$, $g_{0,n}:{\cal X}_{0,n}\times{\cal Y}_{0,n-1}\longmapsto[0,\infty]$ is Borel measurable, and $\overleftarrow{\cal{P}}_{0,n}(P)$ non-empty. In the absence of any power constraints the set of input conditional distributions is ${\cal Q}^{\bf C1}({\cal X}_{0,n};{\cal Y}_{0,n-1})$.\\
The finite horizon maximization of directed information over $\overleftarrow{\cal{P}}_{0,n}(P)$ or ${\cal Q}^{\bf C1}({\cal X}_{0,n};{\cal Y}_{0,n-1})$ (e.g., with or without power constraints) is defined by
\begin{align}
C^{f}_{0,n}\tri\sup_{\substack{{\overleftarrow{P}}_{0,n}(\cdot|y^{n-1})\in\overleftarrow{\cal{P}}_{0,n}(P)\\
~{or}~{\cal Q}^{\bf C1}({\cal X}_{0,n};{\cal Y}_{0,n-1})}}\mathbb{I}_{X^n\rightarrow{Y^n}}({\overleftarrow{P}}_{0,n},{\overrightarrow{Q}}_{0,n})\label{equation44}
\end{align}
The next theorem establishes existence of the maximizer.
\begin{theorem}\label{existence_capacity}
Suppose the assumptions of Theorem~\ref{weak_convergence}, {\bf Part A} are satisfied.
\begin{itemize}
\item[1)] The set ${\cal Q}^{\bf C1}({\cal X}_{0,n};{\cal Y}_{0,n-1})$ is  compact.
\item[2)] Suppose $g_{0,n}:{\cal X}_{0,n}\times{\cal Y}_{0,n-1}\longmapsto[0,\infty]$ is measurable and continuous in $(x^n,y^{n-1})\in{\cal X}_{0,n}\times{\cal Y}_{0,n-1}$. Then the set $\overleftarrow{\cal{P}}_{0,n}(P)$ is a closed subset of ${\cal Q}^{\bf C1}({\cal X}_{0,n};{\cal Y}_{0,n-1})$.
\item[3)] If in addition the assumptions of Theorem~\ref{upper_semicontinuity} are satisfied (here the assumption on ${\cal Q}^{\bf C1}({\cal X}_{0,n};{\cal Y}_{0,n-1})$ is satisfied by 1) and 2) ) then $C_{0,n}^{f}$ has a maximum in ${\cal Q}^{\bf C1}({\cal X}_{0,n};{\cal Y}_{0,n-1})$ (without constraints) or in $\overleftarrow{\cal{P}}_{0,n}(P)$ (with power constraints).
\end{itemize}
\end{theorem}
{\IEEEproof} 1) Utilize the fact that probability measures on compact Polish spaces are compact. 2) Utilize the fact that closed subset of weakly compact set is compact. 3) Follows from Weierstrass theorem.{\endIEEEproof}
\subsection{Existence of Non-Anticipative Rate Distortion Achieving Distribution}
\par Consider a reconstruction channel ${\overrightarrow{Q}}_{0,n}(\cdot|x^n)\in{\cal Q}^{\bf C2}({\cal Y}_{0,n};{\cal X}_{0,n})$, a fixed source $\mu_{0,n}(dx^n)\in{\cal M}_1({\cal X}_{0,n})$, and define the fidelity constraint by
\begin{align}
&\overrightarrow{Q}_{0,n}(D)\tri\big\{{\overrightarrow{Q}}_{0,n}(\cdot|x^{n})\in{\cal Q}^{\bf C2}({\cal Y}_{0,n};{\cal X}_{0,n}):\nonumber\\
&\int{d}_{0,n}(x^n,y^{n})({\mu}_{0,n}\otimes{\overrightarrow{Q}}_{0,n})(dx^n,dy^n)\leq{D}\big\}\label{equation51}
\end{align}
where $D\geq{0}$, and  for each $n\in\mathbb{N}$, $d_{0,n}:{\cal X}_{0,n}\times{\cal Y}_{0,n}\longmapsto[0,\infty]$ is Borel measurable, and $\overrightarrow{Q}_{0,n}(D)$ is non-empty.\\ 
The finite horizon minimization of directed information over $\overrightarrow{Q}_{0,n}(D)$ is defined by
\begin{align}
R^{c}_{0,n}(D)\tri\inf_{{\overrightarrow{Q}}_{0,n}(\cdot|x^{n})\in\overrightarrow{Q}_{0,n}(D)}\mathbb{I}_{X^n\rightarrow{Y^n}}({\mu}_{0,n},{\overrightarrow{Q}}_{0,n})\label{equation52}
\end{align}
The next theorem establishes existence of the minimizer.
\begin{theorem}\label{existence_rd}
Let ${\cal X}_{0,n}$ be a Polish space and ${\cal Y}_{0,n}$ a compact Polish space. Assume $\forall~h(\cdot){\in}BC({\cal Y}_{0,n})$, the function
\begin{align*}
(x^{n},y^{n-1})\in{\cal X}_{0,n}\times{\cal Y}_{0,n-1}\longmapsto\int_{{\cal Y}_n}h(y)q_n(dy;y^{n-1},x^n)\in\mathbb{R}
\end{align*}
is continuous jointly in $(x^{n},y^{n-1})\in{\cal X}_{0,n}\times{\cal Y}_{0,n-1}$.\\
Then
\begin{itemize}
\item[1)] The set ${\cal Q}^{\bf C2}({\cal Y}_{0,n};{\cal X}_{0,n})$ is compact.
\item[2)] Assume $d_{0,n}:{\cal X}_{0,n}\times{\cal Y}_{0,n}\longmapsto[0,\infty]$ is measurable and continuous on $y^n\in{\cal Y}_{0,n}$. Then $\overrightarrow{Q}_{0,n}(D)$ is a closed subset of ${\cal Q}^{\bf C2}({\cal Y}_{0,n};{\cal X}_{0,n})$.
\item[3)] $R_{0,n}^c(D)$ has a minimum in $\overrightarrow{Q}_{0,n}(D)$.
\end{itemize}
\end{theorem}
{\IEEEproof} Utilize Theorem~\ref{weak_convergence}, {\bf Part A}, and generalize the derivation in \cite{csiszar74} to $(n+1)-$fold convolution measures.
{\endIEEEproof}

\section{Conclusion}
In this paper we have provided a general framework through which the properties of mutual information are extended to directed information on Polish spaces. The existence of extremums to capacity problems with memory and feedback, and to lossy non-anticipative data compression problems are discussed. 

\bibliographystyle{IEEEtran}

\bibliography{photis_references}
\end{document}